\documentclass{ws-procs9x6}

\begin{document}
\title{Pair Production In Near Extremal Charged Black Holes}
\author{Chiang-Mei Chen$^{1a}$, Jia-Rui Sun$^{2b}$ and Fu-Yi Tang$^{1c}$}
\address{$^1$Department of Physics, National Central University, Chungli 320, Taiwan
\\
$^2$Institute of Astronomy and Space Science, Sun Yat-Sen University, Guangzhou 510275, China
\\
$^a$cmchen@phy.ncu.edu.tw, $^b$sunjiarui@sysu.edu.cn, $^c$foue.tang@gmail.com}

\begin{abstract}
We study the spontaneous pair production, including the Schwinger mechanism and the Hawking radiation, of charged scalar and spinor particles from the near horizon region of (near) extremal charged black holes in the probe field limit. The pair production rate and the absorption cross section, as well as the retarded Green's functions of the probe fields are analytically computed. Moreover, the holographic description dual to the pair production is discussed.
\end{abstract}


\bodymatter

\section{Introduction}
The spontaneous pair production occurring in charged black holes mixes two independent processes, the Schwinger mechanism~\cite{Schwinger:1951nm} and the Hawking radiation~\cite{Parikh:1999mf}. Since this effect is expected to happen in the near horizon region, we investigate the particle emissions (scalar and spinor particles) in the spacetime of the near-horizon region of the near extremal Reissner-Nordstr{\"o}m (RN) black hole~\cite{Chen:2012zn, Chen:2014yfa} without back-reactions. The spacetime has the geometric structure AdS$_2 \times S^2$ plus a constant electric field, which allows us to analytically solve the Klein-Gorden (KG) equation and the Dirac equation for the probe scalar and spinor fields, respectively and obtain the exact solutions in terms of the well-known hypergeometric functions. Further, by imposing the particle viewpoint boundary condition on the solutions, the physical quantities: the vacuum persistence amplitude, the mean number of produced pairs, the absorption cross section ratio and the retarded Green's function of the probe fields are gained. In addition, the physical quantities calculated from the gravity side are showed to match well with those results of the scalar and spinor operators in the dual boundary conformal field theories (CFTs) side, based on the RN/CFT correspondence~\cite{Chen:2009ht, Chen:2010bsa, Chen:2010as, Chen:2010yu, Chen:2010ywa, Chen:2011gz}. Moreover, the corresponding thermal interpretation has been discussed~\cite{Kim:2015kna, Kim:2015qma, Kim:2015wda}.

\section{The near horizon near extreme RN black holes}
The geometry of the near horizon near extreme RN black hole has the structure of an AdS$_2 \times S^2$ as~\cite{Chen:2012zn}
\begin{eqnarray} \label{NearHorizon}
ds^2 &=& - \frac{\rho^2 - B^2}{Q^2} d\tau^2 + \frac{Q^2}{\rho^2 - B^2} d\rho^2 + Q^2 d\Omega_2^2,
\nonumber\\
A &=& -\frac{\rho}{Q} d\tau; \quad F = \frac{1}{Q} d\tau \wedge d\rho,
\end{eqnarray}
where $\rho$ is the radial coordinate of the AdS$_2$ section, $B$ labels the deviation from extreme limit and acts as the new horizon radius in this geometry, and $Q$ is the charge of the original RN black hole. Geometrically, the positive curvature of $S^2$ exactly compensates the negative part of the AdS$_2$.

\section{Particle creation}
\subsection{Boundary conditions}
There is a potential barrier due to the electromagnetic and gravitational forces in the near horizon region, the pair production becomes a tunneling process by considering the fluxes in the scattering matrix theory. Two boundary conditions can be imposed (see Fig.~\ref{fig}):

\begin{figure}
\centering
\includegraphics[width=4in]{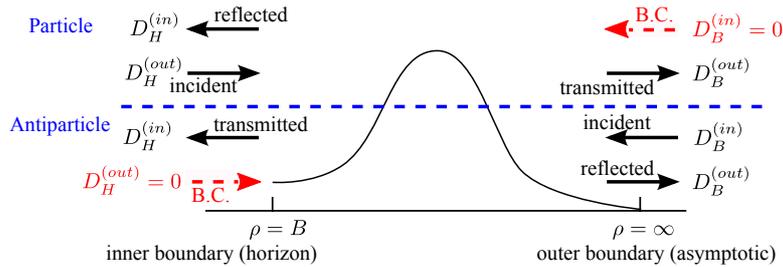}
\caption{Outer boundary condition (upper picture): no incoming flux at asymptotic. Inner boundary condition (lower picture): no outgoing flux at horizon.}
\label{fig}
\end{figure}

The first one is the outer boundary condition (particle viewpoint), which restricts no incoming fluxes at the asymptotic outer boundary. In the St\"uckelberg-Feynman picture, the outgoing (transmitted) flux at the asymptotic represents the spontaneous produced \textit{particle}, the outgoing (incident) flux at the horizon represents the total particles created by vacuum fluctuations, and the incoming (reflected) flux represents the re-annihilated.

The second one is the inner boundary condition (antiparticle viewpoint) that restricts no outgoing fluxes at the inner boundary, which means that the incoming (transmitted) flux at the horizon represents the spontaneous produced \textit{antiparticle}, the incoming (incident) and the outgoing (reflected) fluxes at the asymptotic represents the total created antiparticles and the re-annihilated. Since the particle and anti-particle are always produced in pair due to the charge conservation and/or the energy-momentum conservation, these two boundary conditions are actually equivalent~\cite{Chen:2012zn}.

\subsection{Physical quantities}
For bosonic and fermionic particles, there are two versions of flux conservations~\cite{Kim:2003qp}
\begin{eqnarray}
\text{bosons}: && \quad |D_{\text{incident}}| = |D_{\text{reflected}}| + |D_{\text{transmitted}}|,
\nonumber\\
\text{fermions}: && \quad |D_{\text{reflected}}| = |D_{\text{incident}}| + |D_{\text{transmitted}}|,
\end{eqnarray}
and two versions of Bogoliubov relations
\begin{equation}
\text{bosons}: \quad |\mathcal{A}|^2 - |\mathcal{B}|^2 = 1, \qquad \text{fermions}: \quad |\mathcal{A}|^2 + |\mathcal{B}|^2 = 1.
\end{equation}
where the vacuum persistence amplitude $|\mathcal{A}|^2$ and the mean number of produced pairs $|\mathcal{B}|^2$ are given by the ratio of the flux components in the Coulomb gauge
\begin{equation} \label{Bogoliubov}
|\mathcal{A}|^2 \equiv \frac{D_{\text{incident}}}{D_{\text{reflected}}}, \quad |\mathcal{B}|^2 \equiv \frac{D_{\text{transmitted}}}{D_{\text{reflected}}}.
\end{equation}
Moreover, from the viewpoint of scattering of an incident flux from the asymptotic boundary, we can define the absorption cross section ratio as
\begin{equation} \label{absorptioncrosssection}
\sigma_{\text{abs}} \equiv \frac{D_{\text{transmitted}}}{D_{\text{incident}}} = \frac{|\mathcal{B}|^2}{|\mathcal{A}|^2}.
\end{equation}

\section{Scalar production}
The EoM of the probe scalar field $\Phi$ with mass $m$ and charge $q$ propagating in the near horizon geometry~(\ref{NearHorizon}) is
\begin{equation}
D_\alpha D^\alpha \Phi - m^2 \Phi = 0,
\end{equation}
where $D_\alpha \equiv \nabla_\alpha - i q A_\alpha$ with $\nabla_\alpha$ being the usual covariant derivative. The corresponding radial flux is
\begin{equation} \label{fluxeq}
D = i \sqrt{-g} g^{\rho\rho} (\Phi D_\rho \Phi^* - \Phi^* D_\rho \Phi).
\end{equation}
Using the ansatz $\Phi = \exp(-i \omega \tau + i n \phi) R(\rho) S(\theta)$, the KG equation can be separated into the standard spherical harmonic equation with eigenvalue $\lambda_\ell = \ell (\ell + 1)$ ($\ell$: integer) and the radial part
\begin{equation} \label{radialequation}
\partial_\rho [ (\rho^2 -B^2) \partial_\rho R ] + \left[ \frac{(q \rho - \omega Q)^2 Q^2}{\rho^2 - B^2} - m^2 Q^2 - \lambda_\ell \right] R = 0,
\end{equation}
which resembles the EoM of a scalar field with effective mass $m_{\text{eff}}^2 = m^2 - q^2 + \lambda_\ell/Q^2$ propagating in an AdS$_2$ geometry with radius $L_{\text{AdS}} = Q$. We find that the existence condition for the pair production is the violation of the Breitenlohner-Freedman (BF) bound~\cite{Breitenlohner:1982bm, Breitenlohner:1982jf} in the AdS$_2$ (or effectively AdS$_3$) spacetime, i.e.
\begin{equation}\label{BFscalar}
(m^2 - q^2) Q^2 + \left( \ell + 1/2 \right)^2 < 0,
\end{equation}
which indicates that the mass of the created particle should be smaller than its charge.

Eq.(\ref{radialequation}) can be exactly solved by the following solution
\begin{eqnarray} \label{scalarsolution}
R(\rho) &=& c_1 (\rho - B)^{(-i/2) (a - \tilde{a})}(\rho + B)^{(i/2)(a + \tilde{a})} F\Big( \frac{1}{2} + i \tilde{a} + i b, \frac{1}{2} + i \tilde{a} - i b;
\nonumber\\
&& 1 - i a + i \tilde{a}; \frac{1}{2} - \frac{\rho}{2 B} \Big) + c_2 (\rho - B)^{(i/2) (a - \tilde{a})} (\rho + B)^{(i/2) (a + \tilde{a})}
\nonumber\\
&& F\Big( \frac{1}{2} + i a + i b, \frac{1}{2} + i a - i b; 1 + i a - i \tilde{a}; \frac{1}{2} - \frac{\rho}{2B} \Big),
\end{eqnarray}
with parameters $a \equiv q Q, \tilde{a} \equiv \omega Q^2/B, b \equiv \sqrt{(q^2 - m^2) Q^2 - (\ell + 1/2)^2}$.
Then the Bogoliubov coefficients and the absorption cross section ratio in particle viewpoint can be obtained as
\begin{eqnarray}\label{ABscalar}
|\mathcal{A}|^2 &=& \frac{|D_\text{H}^{(\text{out})}|}{|D^{(\text{in})}_\text{H}|} = \frac{\cosh(\pi a - \pi b) \cosh(\pi \tilde{a} + \pi b)}{\cosh(\pi a + \pi b) \cosh(\pi \tilde{a} - \pi b)},
\nonumber\\
|\mathcal{B}|^2 &=& \frac{|D_\infty^{(\text{out})}|}{|D^{(\text{in})}_\text{H}|} = \frac{\sinh(2 \pi b) \sinh(\pi \tilde{a} - \pi a)}{\cosh(\pi a + \pi b) \cosh(\pi \tilde{a} - \pi b)},
\nonumber\\
\sigma_{\text{abs}} &=& \frac{|\mathcal{B}|^2}{|\mathcal{A}|^2} = \frac{\sinh(2 \pi b) \sinh(\pi \tilde{a} - \pi a)}{\cosh(\pi a - \pi b) \cosh(\pi \tilde{a} + \pi b)}.
\end{eqnarray}
where $D^{(\text{out})}_H, D^{(\text{in})}_H$ are the outgoing and ingoing fluxes at horizon, respectively and $D^{(\text{out})}_\infty$ is the outgoing flux at asymptotic boundary (results for antiparticle viewpoint can be found in~\cite{Chen:2012zn}). In the extremal limit $B \to 0 \, (\tilde{a} \to \infty)$, the leading term of $|\mathcal{B}|^2$ leads to the Schwinger formula $|\mathcal{B}|^2 \simeq e^{- \pi m^2 Q/q} \simeq e^{- \pi m^2 r_H^2/q Q}$. We can compare the ratio $|\mathcal{B}(B = 0)|^2/|\mathcal{B}(B \neq 0)|^2 \geq 1$, which means the production rate in the extremal limit (Schwinger) is greater than that in the near extremal limit (Schwinger + Hawking). In other words, as the geometry changes from the extremal to near extremal black hole, the increasing gravitational force will reduce the electromagnetic repulsive force for the Schwinger mechanism~\cite{Chen:2012zn}.

\section{Spinor production}
The EoM for charged spinor particles is the Dirac equation
\begin{equation} \label{Diracequation}
\left[ \gamma^a e_a{}^\mu \left( \partial_\mu + \Gamma_\mu - i q A_\mu \right) + m \right] \Psi = 0,
\end{equation}
where $e_a{}^\mu$ is the tetrad and $\Gamma = \Gamma_\mu dx^\mu = \frac{1}{4} \gamma^a \gamma^b \omega_{ab}$ is the spin connection one-form. Choosing the ansatz $\Psi = \exp(-i \omega \tau) [R_+(\rho) \Phi^+_{\kappa, n}(\theta, \phi) + R_-(\rho) \Phi^-_{\kappa, n}(\theta, \phi)]$, where $\Phi^{\pm}_{\kappa, n}$ are the spherical spinors correspond to the eigenvalues, $\kappa = \mp (j + 1/2)$ ($j$: half-integer) and the projection $n (- \ell < n < \ell)$, eq.(\ref{Diracequation}) reduces to two first order coupled equations~\cite{Chen:2014yfa}
\begin{equation}\label{1stcoupled}
\frac{\sqrt{\rho^2 - B^2}}{Q} \partial_\rho \mathcal{R}_\pm \mp i \frac{\omega Q - q \rho}{\sqrt{\rho^2 - B^2}} \mathcal{R}_\pm + \left( \frac{\kappa}{Q} \pm i m \right) \mathcal{R}_\mp = 0,
\end{equation}
where $\mathcal{R}_\pm = R_+ \pm R_-$. Using the new coordinate $z = (\rho + B)/2 B$ and two re-scaled functions $\tilde{\mathcal{R}}_\pm = \Sigma^{\pm 1} \mathcal{R}_\pm$ where $\Sigma = (2B)^{i a} z^{i (\tilde{a} + a)/2} (z - 1)^{-i (\tilde{a} - a)/2}$ with the parameters $a \equiv q Q, \tilde{a} \equiv \omega Q^2/B, \bar{b} \equiv \sqrt{(q^2 - m^2) Q^2 - (j + 1/2)^2}$, eq.(\ref{1stcoupled}) can be further reduced to the hypergeometric differential equations
\begin{equation} \label{spinorsolution}
z (1 - z) \partial_z^2 \tilde{\mathcal{R}}_\pm + \left[ \gamma_\pm - (\alpha_\pm + \beta_\pm + 1) z \right] \partial_z \tilde{\mathcal{R}}_\pm - \alpha_\pm \beta_\pm \tilde{\mathcal{R}}_\pm = 0,
\end{equation}
with $\alpha_\pm = i (\bar{b} \mp a), \beta_\pm = -i (\bar{b} \pm a)$ and $\gamma_\pm = \frac{1}{2} \mp i (\tilde{a} + a)$. The solutions are
\begin{eqnarray}
\tilde{\mathcal{R}}_\pm &=& C_\pm F\left( i (\bar{b} \mp a), -i (\bar{b} \pm a); \frac{1}{2} \mp i (\tilde{a} + a); z \right)
\\
&& + \bar{C}_\pm z^{\frac{1}{2} \pm i (\tilde{a} + a)} F\left( \frac{1}{2} - i (\bar{b} \mp \tilde{a}), \frac{1}{2} + i (\bar{b} \pm \tilde{a}); \frac{3}{2} \pm i (\tilde{a} + a); z \right). \nonumber
\end{eqnarray}
Also, the condition for the pair production to happen requires the parameter $\bar{b}$ to be real, namely,
\begin{equation}
(m^2 - q^2) Q^2 + \left( j + 1/2 \right)^2 < 0.
\end{equation}
which breaks the BF bound of spinor fields in the AdS$_2$ (or effectively AdS$_3$) spacetime. Moreover, according to the charge and energy conservation, the black holes lose their charge more than mass in the process of pair production, which ensures the cosmic censorship conjecture when the back-reaction is not included.

Likewise, the Bogoliubov coefficients and the absorption cross section ratio in the particle picture are given by
\begin{eqnarray}\label{ABspinor}
|\mathcal{A}|^2 &=& \frac{|D_\text{H}^{(\text{out})}|}{|D^{(\text{in})}_\text{H}|} = \frac{\sinh(\pi a - \pi \bar{b}) \cosh(\pi \tilde{a} + \pi \bar{b})}{\sinh(\pi a + \pi \bar{b}) \cosh(\pi \tilde{a} - \pi \bar{b})},
\nonumber\\
|\mathcal{B}|^2 &=& \frac{|D_\infty^{(\text{out})}|}{|D^{(\text{in})}_\text{H}|} = \frac{\sinh(2 \pi \bar{b}) \cosh(\pi \tilde{a} - \pi a)}{\sinh(\pi a + \pi \bar{b}) \cosh(\pi \tilde{a} - \pi \bar{b})},
\nonumber\\
\sigma_{\text{abs}} &=& \frac{|\mathcal{B}|^2}{|\mathcal{A}|^2} = \frac{\sinh(2 \pi \bar{b}) \sinh(\pi \tilde{a} - \pi a)}{\sinh(\pi a - \pi \bar{b}) \cosh(\pi \tilde{a} + \pi \bar{b})}.
\end{eqnarray}
It can be seem that results in eq.(\ref{ABspinor}) and eq.(\ref{ABscalar}) are simply related via $\sinh(\pi a \pm \pi \bar{b}) \leftrightarrow \cosh(\pi a \pm \pi b)$ and $\cosh(\pi \tilde{a} - \pi a) \leftrightarrow \sinh(\pi \tilde{a} - \pi a)$, except for $\ell $ is integer, while $j$ is half-integer.

\section{Holographic CFT description}
The near horizon geometry of the near extreme RN black hole is dual to a 2d CFT with left- and right-hand central charges and temperatures
\begin{equation}
c_L = c_R = \frac{6 Q^3}{\ell}, \qquad T_L = \frac{\ell}{2\pi Q}, \qquad T_R = \frac{\ell B}{\pi Q^2},
\end{equation}
where $\ell$ is a parameter that can be interpreted as a measure of the U(1) bundle of the background spacetime~\cite{Chen:2009ht, Chen:2010bsa, Chen:2010as, Chen:2010yu, Chen:2010ywa, Chen:2011gz}.

For the bulk scalar and spinor fields, the conformal dimensions of their dual operators are
\begin{eqnarray}\label{cfmdim}
\text{scalar operator}: && \quad h_L = h_R = \frac12 \pm ib,,
\nonumber\\
\text{spinor operator}: && \quad h_R=\frac12\pm i\bar{b},\quad h_L=1\pm i\bar{b}.
\end{eqnarray}
Also note that the first law of thermodynamics of the black hole is identical to that of the dual CFT, i.e.
\begin{equation}
\frac{\delta M}{T_H} - \frac{\Omega_H \delta Q}{T_H} = \frac{\tilde{\omega}_L}{T_L} + \frac{\tilde{\omega}_R}{T_R},
\end{equation}
in which the black hole Hawking temperature and chemical potential are $T_H = \frac{B}{2 \pi Q^2}, \, \Omega_H = A_\tau(B) = - B/Q$. Then
\begin{equation}
\tilde{\omega}_L = - q \ell \quad {\rm and} \quad \tilde{\omega}_R = 2 \omega \ell.
\end{equation}
Thus the absorption cross sections in eqs.(\ref{ABspinor})(\ref{ABscalar}) match with their CFT's results, respectively, only up to some numerical factors. Consequently, the mean number of pairs $| {\cal B} |^2$ also match with the CFT 2-point functions, both for the scalars and fermions, via the relation $| {\cal B} |^2=-\sigma_\mathrm{abs}(b \to -b)$. Furthermore, the retarded Green's functions and quasinormal modes of the bulk probe scalar and spinor fields also consistent with the results $G_R(\omega_L, \omega_R)$ from the dual CFT$_2$ side in which the frequencies take discrete values of the Matsubara frequencies, respectively.

\section{Conclusions}
We study the scalar and spinor pair production for the near extremal RN black hole without back-reaction. The KG equation and Dirac equation are solved in the near horizon region where the geometry is AdS$_2 \times S^2$ and the background electric field is constant in the radial direction. The near horizon region contains causal horizon and dominated electric field which capture both contributions: the Hawking radiation and the Schwinger mechanism. The exact solutions were obtained in terms of the hypergeometric functions. By imposing the particle viewpoint boundary condition, the physical quantities associated the pair production can be derived by the ratio of boundary fluxes. In particular, the expressions of the vacuum persistence amplitude, the mean number of pairs and the absorption section ratio were obtained. The existence condition for the pair production is corresponding to the instability of the probe fields in the AdS$_2$, i.e. violating the BF bound, leads to the black hole losing their charge more than their mass in pair production process. This consequence is in agreement with the cosmic censorship conjecture that a naked singularity cannot be evolved form the complete gravitational collapse when the matter fields satisfy appropriated energy conditions. We also present the holographic dual IR CFT$_2$ description for the pair production process, it would be more interesting to study the asymptotically AdS black holes and further study the dual UV CFT description for the pair production on the AdS boundary.


\section*{Acknowledgments}
We are grateful to Sang Pyo Kim for useful discussions. C.M.C. and F.Y.T. were supported by the Ministry of Science and Technology of the R.O.C. under the grant MOST 102-2112-M-008-015-MY3. J.R.S. was supported by the NSFC under Grant No.~11205058 and the Open Project Program of State Key Laboratory of Theoretical Physics, Institute of Theoretical Physics, Chinese Academy of Sciences, China (No.~Y5KF161CJ1).


\end{document}